\documentclass[aps,prl,twocolumn,showpacs,superscriptaddress,floatfix]{revtex4}
\usepackage{amssymb}
\usepackage{amsmath}
\usepackage{graphicx}
\usepackage{dcolumn}
\usepackage{bm}
\usepackage[toc,page]{appendix}
\usepackage{natbib,hyperref}
\usepackage{color}

\def\be{\begin{equation}}
\def\ee{\end{equation}}
\input{tcilatex}

\begin{document}

\title{Anisotropic magnetoresistance driven by surface spin orbit scattering}

\begin{abstract}
In a bilayer consisting of an insulator (I) and a ferromagnetic metal (FM),
interfacial spin orbit scattering leads to spin mixing of the two conducting
channels of the FM, which results in an unconventional anisotropic
magnetoresistance (AMR). We theoretically investigate the magnetotransport
in such bilayer structures by solving the spinor Boltzmann transport
equation with generalized Fuchs-Sondheimer boundary condition that takes
into account the effect of spin orbit scattering at the interface. We find
that the new AMR exhibits a peculiar angular dependence which can serve as a
genuine experimental signature. We also determine the dependence of the AMR
on film thickness as well as spin polarization of the FM.

\end{abstract}

\date{\today }

\author{Steven S.-L. Zhang}
\email{zhangshule@missouri.edu} \affiliation{Department of Physics
and Astronomy, University of Missouri, Columbia, MO 65211}
\affiliation{Department of Physics, University of Arizona, Tucson,
AZ 85721}
\author{Giovanni Vignale}
\affiliation{Department of Physics and Astronomy, University of
Missouri, Columbia, MO 65211}
\author{Shufeng Zhang}
\affiliation{Department of Physics, University of Arizona, Tucson,
AZ 85721}

%

\maketitle

Anisotropic magnetoresistance (AMR) is a generic magnetotransport
property of ferromagnetic metals. In general, the longitudinal
resistance of a bulk polycrystalline ferromagnetic metal only
depends on the relative orientations of the magnetization vector and
the current, which can be cast in the form \cite{McGuire75IEEE},
\begin{equation}  \label{AMRStandard}
\rho =\rho _{0}+\Delta \rho _{b}(\hat{\mathbf{j}}_{e}\cdot \mathbf{m})^{2}\,
\end{equation}%
where $\hat{\mathbf{j}}_{e}=\mathbf{j}_{e}/j_{e}$ is the unit vector in the
direction of the current density, $\mathbf{m}$ is the unit vector in the
direction of the magnetization, $\rho _{0}$ is the isotropic longitudinal
resistivity and $\Delta \rho _{b}$ quantifies the magnitude of the bulk AMR
effect (typically $\Delta \rho _{b}\sim 1\%\,$for transition metals and
their alloys). The effect has found many practical applications in magnetic
recording and sensor devices~\cite{sensor08book}.

Recently, the AMR effect has also played a key role in measurements of spin
Hall angle \cite{Hoffmann10PRB_SH,hfDing12PRB,Hoffmann10PRL} as well as spin
torque generation~\cite{lqLiu12Science,lqLiu12PRL_SH-switchinng,lqLiu11PRL,
Ralph14Nature} in FM/heavy-metal and FM/topological-insulator (TI) bilayers.
The structural inversion asymmetry of these structures, combined with strong
spin orbit coupling in the non-magnetic layers, generates a large
Rashba-type spin-orbit coupling~\cite%
{aManchonPRB13,Tokatly15PRB,Stiles13PRB_Rashba-Boltzmann}
\begin{equation}  \label{SOCoupling}
\hat{V}_{s.o.}=-\lambda _{c}^{2}V^{\prime }\left( z\right) \boldsymbol{%
\sigma }\cdot \left( \mathbf{\hat{p}}\times \mathbf{z}\right)\,
\end{equation}%
where $\boldsymbol{\sigma }$ is the Pauli spin matrix, $\mathbf{\hat{p}}$ is
the momentum operator, $\lambda _{c}$ is the effective Compton wave length, $%
\mathbf{z}$ is the unit vector normal to the interface, $V(z)$ is the
potential in the vicinity of the interface (which only varies in the $z-$
direction), and $V^{\prime }(z)$ is its derivative, which is large only in
the interfacial region. A natural question arises: will the interfacial spin
orbit interaction alter the AMR in the FM layer? At first glance, one might
think the spin orbit interaction, commuting with the total in-plane
momentum, $p_{x}$ and $p_{y}$, cannot alter the in-plane resistivity.
However, this argument fails in a ferromagnetic metal since the in-plane
momenta of either spin component are not separately conserved, and the
spin-orbit coupling transfers momentum from one spin channel into the other.

\begin{figure}[tbp]
\includegraphics[width=1.0 \columnwidth]{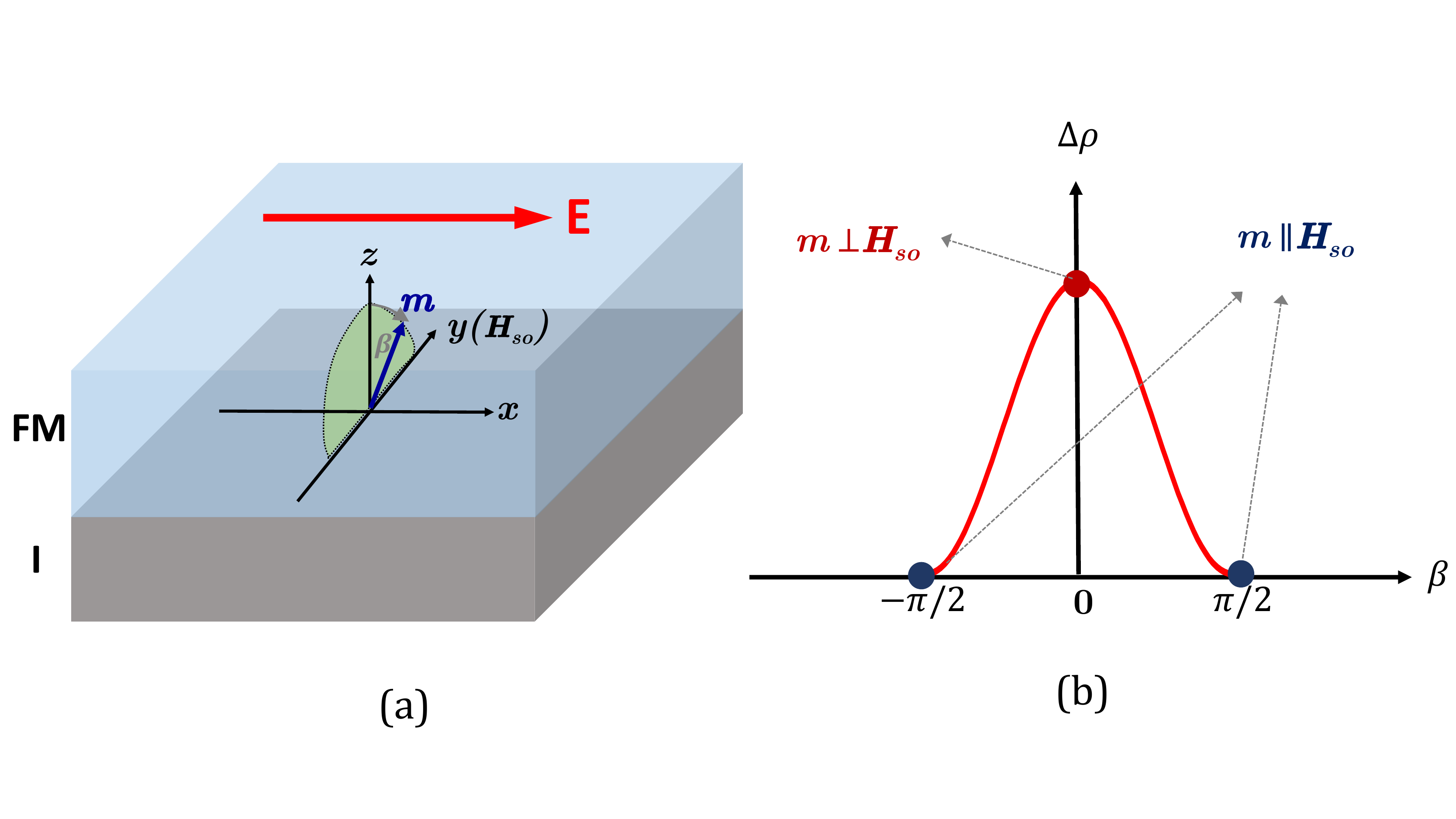}
\caption{Schematics of the transverse AMR effect induced by surface spin
orbit scattering in a FM/I bilayer. The longitudinal resistivity changes
when the magnetization $\mathbf{m}$ is rotated in the plane perpendicular to
the electric field $\mathbf{E}$. Specifically, the resistivity is at a
maximum when $\mathbf{m}$ is perpendicular to the interface (i.e., $\protect%
\beta=0$) and reaches a minimum when $\mathbf{m}$ lies in the plane of the
interface (i.e., $\protect\beta=\pm\protect\pi/2$). This unusual AMR effect
arises from spin mixing of the conducting channels of the FM, which depends
on the relative directions of the magnetization $\mathbf{m}$ and the
effective spin orbital field $\mathbf{H}_{so}\sim \mathbf{z}\times \mathbf{E}
$.}
\label{fig:schematics}
\end{figure}

In this paper, we show that, in the presence of surface spin orbit
scattering, the AMR of a ferromagnet exhibits an angular dependence that is
distinctly different from the conventional one described by Eq.~(1). As
shown in Fig.~1, when the magnetization vector $\mathbf{m}$ is swept in the
plane perpendicular to the applied electric field $\mathbf{E}$, a variation
in the longitudinal resistivity occurs, which has no analogue in Eq.~(\ref%
{AMRStandard}): the resistivity has a maximum when $\mathbf{m}$ is along the
$z-$axis (i.e., normal to the film plane) and reaches a minimum when $%
\mathbf{m}$ is along the $y-$axis (i.e., orthogonal to both $\mathbf{E}$ and
$\mathbf{z}$), even though the angle between the magnetization and the
current does not change. The physical origin of this unconventional angular
dependence lies in the concerted actions of surface spin orbit scattering
and spin asymmetry in the conductivity of the FM, which can be understood
qualitatively within the two-current model~\cite%
{Campbell82book,Fert69JphysC_spin-mixing}. The surface spin orbit scattering
plays a crucial role in mixing the two parallel current channels of majority
and minority spins; moreover, the degree of spin mixing depends on the
relative orientation of the magnetization and the effective magnetic field $%
\mathbf{H}_{so}\sim \mathbf{j}_{e}\times \mathbf{z} $ seen by the electron
spin. Specifically, spin mixing is strong when $\mathbf{m}$ is perpendicular
to $\mathbf{H}_{so}$ while it is weak when $\mathbf{m}$ is aligned with $%
\mathbf{H}_{so}$.  The spin mixing causes, as long as the resistivities of
the two current channels are \emph{not} identical, a redistribution of
current by decreasing the resistivity of the channel with higher resistivity
and increasing the resistivity of the channel with lower resistivity: this
results in an overall increase of the total resistivity~\cite%
{Campbell82book,Fert69JphysC_spin-mixing}. The largest resistivity therefore
coincides with the largest degree of spin mixing, which occurs when $\mathbf{%
m}$ is perpendicular to $\mathbf{H}_{so}$.

In the remainder of this paper we present a quantitative theory of the
spin-orbit driven AMR in a FM/I bilayer with strong interfacial spin orbit
coupling. At variance with previous studies~\cite%
{Saitoh13PRL_SH-MR,Bauer13PRB_SH-MR,slZhang14JAP_AMR,jXiao14PRB_AMR,Chien14PRL_AMR}
we exclude from our consideration heavy metals with strong spin-orbit
coupling: in fact, we assume that the spin-orbit coupling is negligible in
the bulk of the ferromagnet, so as to avoid contamination from the
conventional bulk AMR effect.

Our theoretical analysis is based on the spinor form of the semiclassical
Boltzmann equation: the non-equilibrium distribution function, $\hat{f}(%
\mathbf{k,}z)$, is a $2\times 2$ matrix in spin space~\cite%
{jwZhang04PRL,dyXing97PRB}. For simplicity, we assume same Fermi wave vector
$k_{F}$ but different relaxation times for majority and minority electrons,
just as in the seminal paper by Valet and Fert in calculating the CPP-GMR~%
\cite{Valet93PRB}. Such simplification is justified since the essence of our
effect is in the difference of the relaxation times (and hence the
resistivities) of majority and minority spins, while the exchange splitting
is responsible for spin dephasing of the transverse spin component~\cite%
{Stiles02PRB,jwZhang04PRL}.

The equation of motion for $\hat{f}(\mathbf{k,}z)$ in the steady state is
\begin{equation}
v_{z}\frac{\partial \hat{f}(\mathbf{k,}z)}{\partial z}-eEv_{x}\left( \frac{
\partial f_{0}(k)}{\partial \varepsilon _{k}}\right) \hat{I}=-\frac{1}{2}
\left\{ \hat{\tau}^{-1},\delta \hat{f}(\mathbf{k,}z) \right\}
\label{spinor-Boltzmann}
\end{equation}
where $f_{0}(k)$ is the equilibrium distribution function, $\hat{\tau}%
^{-1}=\left( \tau ^{\prime }\right) ^{-1}\left( \hat{I}-p\boldsymbol{\sigma }%
\cdot \mathbf{m}\right) $ is the spin dependent relaxation rate with $\left(
\tau ^{\prime }\right) ^{-1}=\left[\left( \tau _{\uparrow}\right)
^{-1}+\left( \tau _{\downarrow}\right) ^{-1}\right]/2$ and $p\equiv \left[%
\left( \tau _{\downarrow}\right) ^{-1}-\left( \tau _{\uparrow}\right) ^{-1}%
\right] /\left[\left( \tau _{\uparrow}\right) ^{-1}+\left( \tau
_{\downarrow}\right) ^{-1}\right]$ being, respectively, the average momentum
relaxation rate and the spin asymmetry in resistivity, $\tau_\uparrow$ and $%
\tau_\downarrow$ being the momentum lifetimes of the two spin channels, and $%
\left\{ \text{ },\right\}$ standing for an anticommutator. Notice that in
the collision term of Eq.~(\ref{spinor-Boltzmann}) we require that the
difference
\begin{equation}
\delta \hat f(\mathbf{k},z)\equiv \hat f(\mathbf{k},z)-\hat{f}_{l.e.}(k,z)
\end{equation}
between the non-equilibrium distribution and a \textit{local equilibrium
distribution}~\cite{Haug96book,gVignale08book,Bulter01book}
\begin{equation}
\hat{f}_{l.e.}(k,z)=f_{0}(k)\hat{I}+\frac{\partial f_{0}}{\partial
\varepsilon _{k}}\left[ \mu _{0}\left( z\right) \hat{I}+\boldsymbol{\sigma }%
\cdot \boldsymbol{\mu }_{s}\left( z\right) \right]
\end{equation}
tend to zero for long times. The parameters $\mu_0(z)$ and $\boldsymbol{\mu}%
_s(z)$ of the local distribution are fixed in such a way that the condition
\begin{equation}
\int d^{3}\mathbf{k}\left[ \hat{f}(\mathbf{k,}z)-\hat{f}_{l.e.}(k,z)\right]
=0  \label{continuity}
\end{equation}
is self-consistently satisfied. By doing this, we satisfy the physical
requirements of particle and spin conservation in the collision processes,
as well as the continuity equations that go with them.

To solve the Boltzmann equation, we further separate the distribution
function into an equilibrium part and a small non-equilibrium perturbation,
i.e.,
\begin{equation}
\hat{f}(\mathbf{k,}z)=f_{0}(k)\hat{I}\mathbf{+}\frac{\partial f_{0}}{%
\partial \varepsilon _{k}}\left[ g\left( \mathbf{k,}z\right) \hat{I}+\mathbf{%
h}(\mathbf{k,}z)\cdot \boldsymbol{\sigma }\right]  \label{sol_linear}
\end{equation}%
where $g\left( \mathbf{k,}z\right) \hat{I}$ and $\mathbf{h}(\mathbf{k,}%
z)\cdot \boldsymbol{\sigma }$ are the spin-independent and spin-dependent
components of the non-equilibrium distribution. By inserting Eq.~(\ref%
{sol_linear}) into Eq.~(\ref{spinor-Boltzmann}), we obtain a set of coupled
equations for the scalar and vector parts of the distribution function

\begin{equation}
v_{z}\frac{\partial g}{\partial z}-eE_{x}v_{x}=-\frac{g\mathbf{-}\mu
_{0}-p\left( h_{\parallel }-\mu _{s\parallel }\right) }{\tau ^{\prime }}
\label{v_zPf1/Pz}
\end{equation}

\begin{equation}
v_{z}\frac{\partial h_{\parallel }}{\partial z}=-\frac{h_{\parallel }-\mu
_{s\parallel }-p\left( g-\mu _{0}\right) }{\tau ^{\prime }}  \label{vzPg1Pz}
\end{equation}%
and
\begin{equation}
v_{z}\frac{\partial \mathbf{h}_{\perp }}{\partial z}=-\frac{\mathbf{h}%
_{\perp }-\boldsymbol{\mu }_{s\perp }}{\tau ^{\prime }}  \label{vzPh-perp/Pz}
\end{equation}

\noindent where $h_{\parallel }=\mathbf{m\cdot h}(\mathbf{k,}z)$, $\mu
_{s\parallel }=\mathbf{m\cdot }\boldsymbol{\mu }_{s}\left( z\right) $, $%
\mathbf{h}_{\perp }=\left( \mathbf{m\times h}\right) \mathbf{\times m}$ and $%
\boldsymbol{\mu }_{s\perp }=\left( \mathbf{m\times }\boldsymbol{\mu }%
_{s}\right) \mathbf{\times m}$. The equations (\ref{v_zPf1/Pz})-(\ref%
{vzPh-perp/Pz}) have the general solutions
\begin{widetext}
\begin{equation}
g^{\pm }\left( \mathbf{k,}z\right) =e\tau v_{x}E_{x}+A_{\mathbf{k}}^{\pm
}e^{\mp \frac{\left( 1-p\right) z}{\left\vert v_{z}\right\vert \tau ^{\prime
}}}+B_{\mathbf{k}}^{\pm }e^{\mp \frac{\left( 1+p\right) z}{\left\vert
v_{z}\right\vert \tau ^{\prime }}}+\sum_{\alpha }\int_{0}^{z}dt\left[ \mu
_{0}\left( t\right) +\alpha \mu _{s\parallel }\left( t\right) \right] \frac{%
\partial }{\partial t}e^{\mp \frac{\left( 1-\alpha p\right) \left(
z-t\right) }{\left\vert v_{z}\right\vert \tau ^{\prime }}}  \label{g^+-}\,,
\end{equation}%
\begin{equation}
h_{\parallel }^{\pm }\left( \mathbf{k,}z\right) =pe\tau v_{x}E_{x}+A_{%
\mathbf{k}}^{\pm }e^{\mp \frac{\left( 1-p\right) z}{\left\vert
v_{z}\right\vert \tau ^{\prime }}}-B_{\mathbf{k}}^{\pm }e^{\mp \frac{\left(
1+p\right) z}{\left\vert v_{z}\right\vert \tau ^{\prime }}}+\sum_{\alpha
}\alpha \int_{0}^{z}dt\left[ \mu _{0}\left( t\right) +\alpha \mu
_{s\parallel }\left( t\right) \right] \frac{\partial }{\partial t}e^{\mp
\frac{\left( 1-\alpha p\right) \left( z-t\right) }{\left\vert
v_{z}\right\vert \tau ^{\prime }}}\,,
\end{equation}%
\end{widetext}
and
\begin{equation}
\mathbf{h}_{\perp }^{\pm }\left( \mathbf{k,}z\right) \mathbf{=C_{\mathbf{k}%
}^{\pm }}e^{\mp \frac{z}{\left\vert v_{z}\right\vert \tau ^{\prime }}%
}+\int_{0}^{z}dt\boldsymbol{\mu }_{s\perp }\left( t\right) \frac{\partial }{%
\partial t}e^{\mp \frac{\left( z-t\right) }{\left\vert v_{z}\right\vert \tau
^{\prime }}} \,,  \label{h^+-}
\end{equation}
where the superscript $+$ labels the solution for $v_{z}>0$ and the
subscript $-$ for $v_{z}<0$. The sum over $\alpha$ runs over the values $%
\alpha =\pm 1$. The four unknown constants $A_{\mathbf{k}}$, $B_{\mathbf{k}}$%
, and $\mathbf{C_{\mathbf{k}}}$ (where $\mathbf{C_{\mathbf{k}}}$ is a vector
orthogonal to $\mathbf{m}$, hence with only two components) will now be
determined from the boundary conditions.

Up to this point, the interfacial spin-orbit interaction has not appeared in
our calculations. In particular, the collision term in Eq.~(\ref%
{spinor-Boltzmann}) did not contain it, and therefore conserved spin. The
spin-orbit coupling appears in the boundary condition that connects the
distribution function for electrons impinging on the interface (label $-$)
to the distribution function for electrons that are scattered off the
interface (label $+$). Specifically, in order to take into account the
rotation of spin upon scattering off the interface with the potential
\begin{equation}
\hat{V}_{scat.}=V_{b}\Theta \left( -z\right) -\left( V_{b}\lambda
_{c}^{2}\right) \delta \left( z\right) \boldsymbol{\sigma }\cdot \left(
\mathbf{\hat{p}}\times \mathbf{z}\right)
\end{equation}%
(where $V_{b}$ is the barrier height of the insulator, $\Theta \left(
z\right) $ is the unit step function, $\mathbf{z}$ is the unit vector normal
to the interface and the delta function confines the SO coupling to the
interface at $z=0$), we introduce the following spinor generalization of the
Fuchs-Sondheimer boundary condition~\cite%
{Fuchs38,Sondheimer50PR,Sondheimer48PRS}:%
\begin{equation}
\hat{f}^{+}(\mathbf{k};z=0)=s_{I}\hat{R}^{\dag }\hat{f}^{-}(\mathbf{k;}z=0)%
\hat{R}+\left( 1-s_{I}\right) f_{0}\hat{I}  \label{BC-z=0}
\end{equation}%
where the superscripts $+$ and $-$ correspond to the distribution functions
with $v_{z}>0$ and $v_x<0$ respectively, the parameter $s_{I}$ varies
between 0 and 1, characterizing the fraction of electrons being specularly
reflected \cite{Sondheimer48PRS} ($s_{I}=1$ when the interface is perfectly
smooth and $s_{I}=0$ when the interface is extremely rough) and $\hat{R}$ is
a $2\times 2$ reflection amplitude matrix in spin space which captures the
spin rotation of the reflected electrons. Explicitly,
\begin{equation}
\hat{R}=\frac{\left[ -k_{b}^{2}+\left( \lambda _{c}k_{b}\right) ^{4}q^{2}%
\right] \hat{I}+2i\left( \lambda _{c}k_{b}\right) ^{2}k_{z}\boldsymbol{%
\sigma }\cdot \left( \mathbf{q}\times \mathbf{z}\right) }{\left( \kappa
-ik_{z}\right) ^{2}-\left( \lambda _{c}k_{b}\right) ^{4}q^{2}}  \label{R}
\end{equation}%
where $\kappa \equiv \sqrt{k_{b}^{2}-k_{z}^{2}}$ with $k_{b}\equiv \sqrt{%
2m_{e}^{\ast }V_{b}/\hbar ^{2}}$. The derivation of $\hat R$ is presented in
the appendix. It must be pointed out that the boundary condition described
by Eq.~(\ref{BC-z=0}) neglects the interference between incident and
reflected states. 
However, we neglect this quantum coherence by assuming a rough interface.
This is justified when the characteristic size as well as the correlation
length of the surface roughness is comparable to the Fermi wavelength in
which case the phase coherence is destroyed by surface roughness~\cite%
{Ziman60book}.

For simplicity, we assume spin independent specular reflection only at the
outer surface of the FM layer ($z=d$), i.e.,
\begin{equation}
\hat{f}^{+}(\mathbf{k;}z=d)=\hat{f}^{-}(\mathbf{k;}z=d)
\end{equation}
Neglecting spin-dependent scattering from the other surface simplifies the
calculation without altering any qualitative features of the results. By
inserting Eq.~(\ref{sol_linear}) into the boundary conditions as well as Eq.
(\ref{continuity}) , we can find unique solutions for $\hat{f}(\mathbf{k,}z)$
and the charge current density can be calculated as
\begin{equation}
\mathbf{j}_{e}\left( z\right) =\frac{-e}{\left( 2\pi \right) ^{3}}Tr\int d%
\mathbf{k}\hat{f}(\mathbf{k,}z)\mathbf{v}
\end{equation}

After some algebra, we find the charge current density up to second order in
the spin orbit coupling, i.e., $O\left( \left( \lambda _{c}k_{b}\right)
^{4}\right) $%
\begin{equation}
\mathbf{j}_{e}\left( z\right) =c_{0}E_{x}\left\{ \left[ 1\mathbf{-}\alpha
_{xx}^{(1)}\left( z\right) -\alpha _{xx}^{(2)}\left( z\right) \right]
\mathbf{\hat{x}+}\alpha _{yx}\left( z\right) \mathbf{\hat{y}}\right\}
\end{equation}%
where $c_{0}=e^{2}\tau k_{F}^{3}/3\pi m_{e}^{\ast }$ is Drude conductivity
and two position dependent coefficients read
\begin{equation}
\alpha _{xx}^{(1)}\left( z\right) =\left( 1-s_{I}\right) \sum_{\sigma
}\left( 1+\sigma p\right) F_{p\sigma }\left( z\right)  \label{alpha^(1)_xx}
\end{equation}%
\begin{equation}
\alpha _{xx}^{(2)}\left( z\right) =s_{I}p\left( \lambda _{c}k_{F}\right)
^{4} \left[ 4-\left( m_{x}^{2}+3m_{y}^{2}\right) \right] \sum_{\sigma
}\sigma G_{p\sigma }\left( z\right) \,,  \label{alpha^(2)_xx}
\end{equation}%
and
\begin{equation}
\alpha _{yx}\left( z\right) =s_{I}p\left( \lambda _{c}k_{F}\right)
^{4}m_{x}m_{y}\sum_{\sigma }\sigma G_{p\sigma }\left( z\right) \,,
\label{alpha_yx}
\end{equation}%
where
\begin{equation}
F_{p\sigma }\left( z\right) \equiv \frac{3}{4}\int_{0}^{1}d\xi \frac{(1-\xi
^{2})\cosh \left[ \frac{\left( 1-\sigma p\right) \left( d-z\right) }{\xi
\lambda _{0}\left( 1-p^{2}\right) }\right] }{\exp \left[ \frac{\left(
1-\sigma p\right) d}{\xi \lambda _{0}\left( 1-p^{2}\right) }\right]
-s_{I}\exp \left[ -\frac{\left( 1-\sigma p\right) d}{\xi \lambda _{0}\left(
1-p^{2}\right) }\right] }  \label{F_psigma}
\end{equation}%
and
\begin{equation}
G_{p\sigma }\left( z\right) \equiv \frac{3}{2}\int_{0}^{1}d\xi \frac{\xi
(1-\xi ^{2})^{3/2}\cosh \left[ \frac{\left( 1-\sigma p\right) \left(
d-z\right) }{\xi \lambda _{0}\left( 1-p^{2}\right) }\right] }{\exp \left[
\frac{\left( 1-\sigma p\right) d}{\xi \lambda _{0}\left( 1-p^{2}\right) }%
\right] -s_{I}\exp \left[ -\frac{\left( 1-\sigma p\right) d}{\xi \lambda
_{0}\left( 1-p^{2}\right) }\right] }\,,  \label{G_psigma}
\end{equation}%
and $\lambda _{0}\equiv v_{F}(\tau_\uparrow+\tau_\downarrow)/2$ is the
average electron mean free path. The first term, $\alpha _{xx}^{(1)}$, is
independent of the magnetization direction: it is the resistivity due to
interfacial roughness~\cite{Ziman60book}. The third term, $\alpha _{yx}$,
corresponds to the well-known planar Hall effect~\cite%
{McGuire75IEEE,Campbell82book}.

The second term, $\alpha _{xx}^{(2)}$, describes the new AMR effect. We note
that this effect is of second order in the spin-orbit coupling and vanishes
when the spin polarization $p=0$. The experimentally relevant quantity is
the spatially averaged longitudinal resistivity, which is obtained from the
formula $\rho_{xx}^{-1}\left( d\right) =\left( 1/d\right)
\int_{0}^{d}dz~j_{e,x}\left( z\right) /E_{x}$. As we discussed earlier, the
bulk spin-orbit coupling (not included in our calculation) produces a
conventional AMR with angular dependence shown in Eq.~(1). Therefore in
general, the longitudinal resistivity of the FM thin film should take the
form
\begin{equation}
\rho _{xx}\left( d\right) =\rho _{0}+\Delta \rho _{so}^{\left( b+s\right)
}m_{x}^{2}-\Delta \rho _{so}^{\left( s\right) }m_{y}^{2}  \label{ru_xx}
\end{equation}%
where the first term on the rhs of Eq.~(\ref{ru_xx}) is the isotropic
resistivity and the second term is the AMR with conventional angular
dependence of $(\hat{\mathbf{j}}_{e}\cdot \mathbf{m})^{2}$ to which both
bulk and surface spin orbit coupling may contribute. The most interesting
term is the third term which is solely due to the surface spin orbit
scattering and can be distinguished from the bulk AMR based on the different
angular dependence.

\begin{figure}[tbp]
\includegraphics[width=0.9\columnwidth]{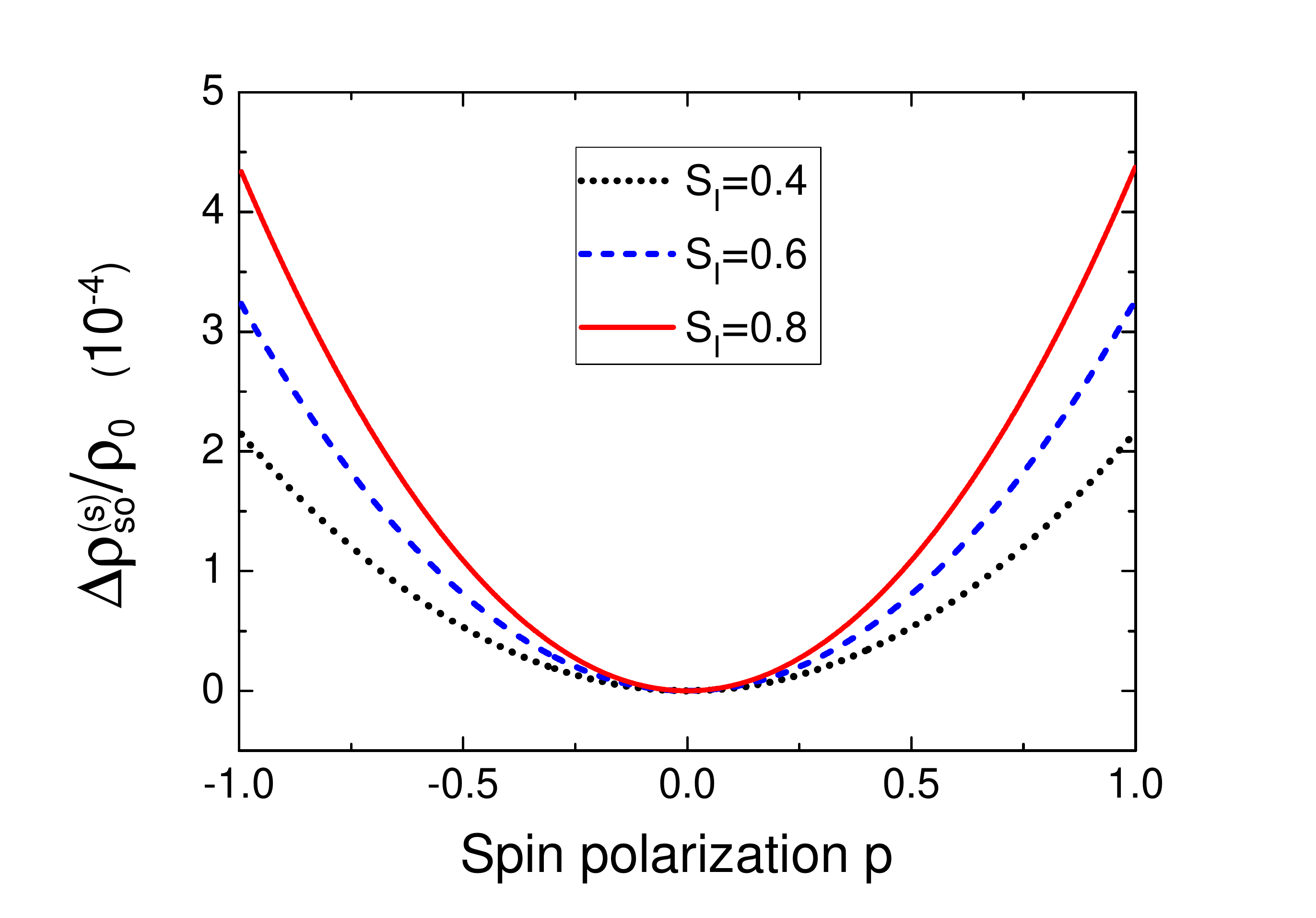}
\caption{The transverse AMR $\Delta \protect\rho _{so}^{\left( s\right)}$ as
a function of the spin polarization $p$ for several different $s_{I}$.
Parameters: $\protect\lambda_c k_F=0.05$, $d/\protect\lambda_0=1$.}
\label{fig:polarization}
\end{figure}
\begin{figure}[tbp]
\includegraphics[width=0.75\columnwidth]{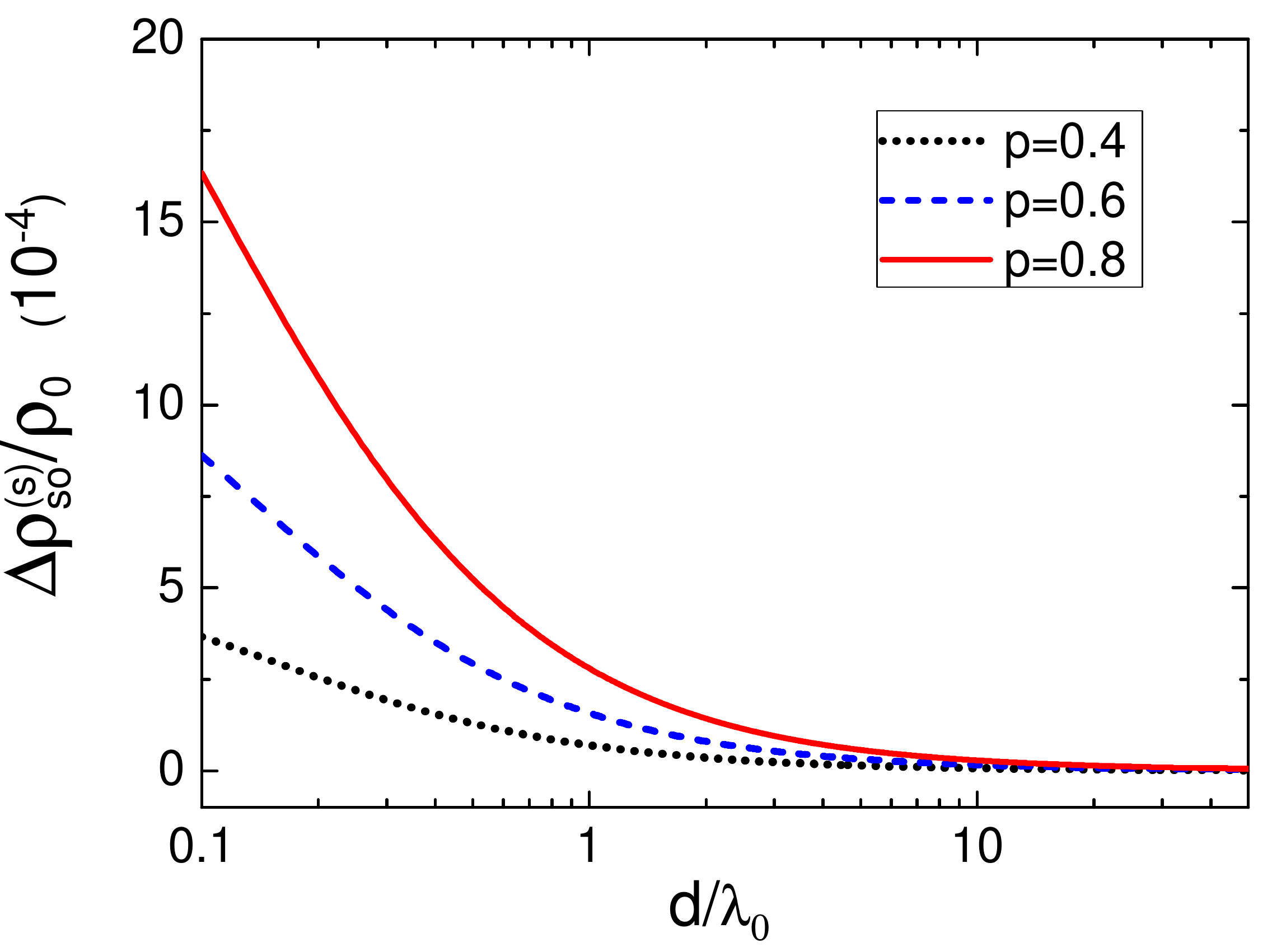}
\caption{The transverse AMR $\Delta \protect\rho _{so}^{\left( s\right) }$
as a function of $d/\protect\lambda _{0}$ for several different $p$.
Parameters: $\protect\lambda _{c}k_{F}=0.05$ and $s_{I}=0.8$.}
\label{fig:thick}
\end{figure}
In Fig.~2, we show $\Delta \rho _{so}^{\left( s\right) }$ (normalized by $%
\rho _{0}$) as a function of the spin polarization $p$. This figure delivers
two main messages. First, $\Delta \rho _{so}^{\left( s\right) }$ is positive
when $p\neq 0$. This confirms the angular dependence of the AMR we sketched
in Fig. 1(b). The second message is that $\Delta \rho _{so}^{\left( s\right)
}$ is an even function of the spin polarization $p$. This is consistent with
our two-current model: the spin mixing resistivity only relies on the
absolute value of the difference between the two conduction channels but not
on the sign of that difference.

In Fig.~3, we show the thickness dependence of the $\Delta \rho
_{so}^{\left( s\right) }$ for several values of spin polarization. When the
FM layer thickness is much larger than the mean free path, i.e., $d\gg
\lambda _{0}$, $\Delta \rho _{so}^{\left( s\right) }$ exhibits a standard $%
1/d$ thickness dependence as can be analytically worked out via Eq.~(\ref%
{G_psigma}).

Lastly, let us consider the choice of materials for the observation of our
predicted AMR. In order to obtain a sizable transverse AMR, it is essential
to have a FM/I bilayer with a large difference between the conductivities of
majority and minority spin carriers in the ferromagnetic metal, and a strong
spin-orbit interaction at the FM/I interface. A very promising system in
this respect is Py grown on top of a TI such as Bi$_{2}$Se$_{3}$, Bi$_{1.5}$%
Sb$_{0.5}$Te$_{1.7}$Se$_{1.3}$ or Sn-doped Bi$_{2}$Te$_{1.7}$Se$_{1.3}$~\cite%
{Taskin12PRB,Taskin11PRL,Taskin11PRB}. Recently, large spin transfer torque~%
\cite{Ralph14Nature} and spin-charge conversion~\cite{Saitoh14PRL_TI}
effects were observed in these FM/TI bilayers, indicating the presence of
strong spin orbit coupling at the interface.
Non-topological-oxide/ferromagnetic interfaces may also provide large
spin-orbit interaction, as implied by tunneling AMR studies in Fe/MgO/Fe
junctions~\cite{Parkin07PRL,Tsymbal07PRL} and by magnetic anisotropy
analysis at AlOx/Co interface~\cite{Parkin07PRL, Manchon08JAP}. As a final
point, we note that the surface spin orbit scattering mechanism should, in
principle, also contribute to the transverse AMR that was previously found
in Pt/Co/Pt~\cite{Kobs11PRL} and Py/YIG~\cite{Chien13PRB_proximity} layered
systems.

This work was supported by NSF Grants DMR-1406568 (S.S.-L.Z and G.V.) and
ECCS-1404542 (S.S.-L.Z and S. Z.).

\appendix

\section{Appendix: Spinor form reflection amplitude in the presence of
interface Rashba spin orbit coupling}

In this appendix, we derive the spinor form reflection amplitude
given by Eq.~(16) in the main text. First, we write down the
following piece-wise scattering wave functions corresponding to the
interfacial potential described by Eq.~(14) in the main text

\begin{equation}
\psi _{\mathbf{k}}\left( \mathbf{r}_{\parallel }\mathbf{,}z>0\right) =\frac{1%
}{\sqrt{2}}e^{-ik_{z}z}e^{i\mathbf{q\cdot r}_{\parallel }}\chi +\frac{1}{%
\sqrt{2}}e^{ik_{z}z}e^{i\mathbf{q\cdot r}_{\parallel }}\hat{R}\chi
\tag{A1}
\end{equation}%
and

\begin{equation}
\psi _{\mathbf{k}}\left( \mathbf{r,}z<0\right)
=\frac{1}{\sqrt{2}}e^{\kappa z}e^{i\mathbf{q\cdot r}_{\parallel
}}\hat{T}\chi   \tag{A2}
\end{equation}%
where $\mathbf{r}_{\parallel }$ is the in-plane position vector,
$\mathbf{q}$ and $k_{z}$ are the in-plane and perpendicular-to-plane
wave vectors
respectively, $\kappa =\sqrt{k_{b}^{2}-k_{z}^{2}}$with $k_{b}\equiv \sqrt{%
2m_{e}^{\ast }V_{b}/\hbar ^{2}}$, $\hat{R}$ and $\hat{T}$ are the
$2\times 2$ reflection and transmission amplitude matrices in spin
space, and $\chi $ is an arbitrary spinor.

Now we are ready to determine $\hat{R}$ and $\hat{T}$ matrices by
the following quantum mechanical matching conditions

\begin{equation}
\psi _{\mathbf{k}}\left( \mathbf{r}_{\parallel
}\mathbf{,}0^{+}\right) =\psi _{\mathbf{k}}\left(
\mathbf{r}_{\parallel }\mathbf{,}0^{-}\right)   \tag{A3}
\end{equation}%
and
\begin{equation}
\psi _{\mathbf{k}}^{\prime }\left( \mathbf{r}_{\parallel }\mathbf{,}%
0^{+}\right) -\psi _{\mathbf{k}}^{\prime }\left( \mathbf{r}_{\parallel }%
\mathbf{,}0^{-}\right) =\left[ \left( k_{b}\lambda _{c}\right) ^{2}%
\boldsymbol{\sigma }\cdot \left( \mathbf{\hat{p}}\times \mathbf{z}\right) %
\right] \psi _{\mathbf{k}}\left( \mathbf{r}_{\parallel
}\mathbf{,}0\right) \tag{A4}
\end{equation}%
By placing the scattering wave functions into the above two
equations, we find
\begin{equation}
\left( \hat{I}+\hat{R}\right) \chi =\hat{T}\chi   \tag{A5}
\end{equation}%
and
\begin{equation}
\left( -\kappa \hat{T}-ik_{z}\hat{I}+ik_{z}\hat{R}\right) \chi
=\left[
\left( k_{b}\lambda _{c}\right) ^{2}\boldsymbol{\sigma }\cdot \left( \mathbf{%
q}\times \mathbf{z}\right) \right] \hat{T}\chi   \tag{A6}
\end{equation}%
Combining the two equations, we find an equation for $\hat{R}$ only,
\begin{widetext}
\begin{equation}
\left\{ \left[ \left( \kappa -ik_{z}\right) \hat{I}+\left(
k_{b}\lambda
_{c}\right) ^{2}\boldsymbol{\sigma }\cdot \left( \mathbf{q}\times \mathbf{z}%
\right) \right] \hat{R}+\left( ik_{z}+\kappa \right) \hat{I}+\left(
k_{b}\lambda _{c}\right) ^{2}\boldsymbol{\sigma }\cdot \left( \mathbf{q}%
\times \mathbf{z}\right) \right\} \chi =0  \tag{A7}
\end{equation}%
For any $\chi $, the equation is valid if
\begin{equation}
\left[ \left( \kappa -ik_{z}\right) \hat{I}+\left( k_{b}\lambda
_{c}\right)
^{2}\boldsymbol{\sigma }\cdot \left( \mathbf{q}\times \mathbf{z}\right) %
\right] \hat{R}+\left( ik_{z}+\kappa \right) \hat{I}+\left(
k_{b}\lambda
_{c}\right) ^{2}\boldsymbol{\sigma }\cdot \left( \mathbf{q}\times \mathbf{z}%
\right) =0  \tag{A8}  \label{Eq.R=0}
\end{equation}%
\end{widetext}
From Eq.~(\ref{Eq.R=0}), we find the reflection amplitude matrix
\begin{equation}
\hat{R}=\frac{\left[ -k_{b}^{2}+\left( k_{b}\lambda _{c}\right) ^{4}q^{2}%
\right] \hat{I}+2i\left( k_{b}\lambda _{c}\right) ^{2}k_{z}\boldsymbol{%
\sigma }\cdot \left( \mathbf{q}\times \mathbf{z}\right) }{\left(
ik_{z}-\kappa \right) ^{2}-\left( k_{b}\lambda _{c}\right)
^{4}q^{2}} \tag{A9}  \label{R^hat}
\end{equation}%
. Note that $\hat{R}$ is a unitary matrix as can be easily check via Eq. (%
\ref{R^hat}). This unitarity is required by flux conservation.

\bibliographystyle{my-asp-style}
\bibliography{20150325_Rashba-AMR}

\end{document}